# Non-benzenoid high-spin polycyclic hydrocarbons generated by atom manipulation

Shantanu Mishra,[†*] Shadi Fatayer,[†] Saleta Fernández,[‡] Katharina Kaiser,[†] Diego Peña[‡*] and Leo Gross[†*]

[†]IBM Research–Zurich, 8803 Rüschlikon, Switzerland.

[‡]Centro Singular de Investigación en Química Biolóxica e Materiais Moleculares (CiQUS) and Departamento de Química Orgánica, Universidade de Santiago de Compostela, 15782 Santiago de Compostela, Spain

*Keywords: scanning tunneling microscopy, atomic force microscopy, on-surface synthesis, open-shell molecules, density functional theory, non-benzenoid hydrocarbons*

**ABSTRACT:** We report the on-surface synthesis of a non-benzenoid triradical through dehydrogenation of truxene ($C_{27}H_{18}$) on coinage metal and insulator surfaces. Voltage pulses applied via the tip of a combined scanning tunneling microscope/atomic force microscope were used to cleave individual C–H bonds in truxene. The resultant final product truxene-5,10,15-triyl (**1**) was characterized at the single-molecule scale using a combination of atomic force microscopy, scanning tunneling microscopy and scanning tunneling spectroscopy. Our analyses show that **1** retains its open-shell quartet ground state, predicted by density functional theory, on a two monolayer-thick NaCl layer on a Cu(111) surface. We image the frontier orbital densities of **1** and confirm that they correspond to spin-split singly occupied molecular orbitals. Through our synthetic strategy, we also isolate two reactive intermediates toward the synthesis of **1** – derivatives of fluorenyl radical and indeno[1,2–*a*]fluorene, with predicted spin-doublet and spin-triplet ground states, respectively. Our results should have bearings on the synthesis of non-benzenoid high-spin polycyclic frameworks with magnetism beyond Lieb's theorem.

## INTRODUCTION

Polycyclic conjugated hydrocarbons (PCHs) have emerged as an important molecular platform for investigating structure-property relationships and emergent phenomena in purely organic materials. Given their finite size, the electronic properties of PCHs critically depend on the molecular size, shape, edge structure and the presence of non-benzenoid rings. In particular, by virtue of their molecular topology, certain PCHs may exhibit an open-shell (magnetic) ground state, which makes them highly attractive for applications in molecular electronics, spintronics and non-linear optics.[1–3] The synthetic chemistry of open-shell PCHs has gained unprecedented traction since the advent of on-surface synthesis,[4] where reactive open-shell molecules that are otherwise unstable in the solution phase can be synthesized and stabilized on solid surfaces under ultra-high vacuum conditions. Additionally, molecules synthesized via this approach are amenable to atomic-scale structural and electronic characterization using scanning probe techniques.[5–7]

A prototypical example of open-shell PCHs is the family of [*n*]triangulenes, which are non-Kekulé triangular PCHs containing *n* benzenoid rings along each edge (Fig. 1a). Since the attempt of Clar and Stewart in 1953 to synthesize [3]triangulene,[8] the synthesis of [*n*]triangulenes remained an enduring challenge for almost 70 years.[9,10] In 2017, on-surface generation of [3]triangulene was reported,[11] where its three-fold symmetric structure and spin-polarized frontier states were visualized by means of scanning probe microscopy. Since then, on-surface syntheses of larger triangulene homologues[12–14] and other triangulene-based open-shell PCHs[15–17] have been reported. Magnetism in [*n*]triangulenes arises as a direct consequence of a sublattice imbalance in the bipartite honeycomb lattice, such that it is impossible to pair up all $p_z$ electrons into π-bonds, which thus generates radicals. An intuitive way to predict the magnetic correlation between radicals in PCHs is Lieb's theorem,[18,19] which states that the ground state total spin quantum number $S = (N–N^*)/2$, where $N$ and $N^*$ denote the two interpenetrating triangular sublattices of the honeycomb lattice. For [*n*]triangulenes, $N–N^* = n–1 > 0$, which leads to a high-spin ground state ($S > 0$) with a linear scaling of $S$ with $n$.

Among the multitude of open-shell molecules realized to date both in solution and on surfaces,[20–22] an overwhelming majority are benzenoid systems. However, few examples of non-benzenoid open-shell systems exist in literature,[23–32] despite the unique opportunities they offer in investigating the interplay of aromaticity and magnetism,[27,33,34] and electronic structure tuning of carbon-based nanostructures.[29,35] Moreover, it has been recently proposed that non-benzenoid biradicals with triplet ground states, formed in hydrocarbon flames, play an important role in incipient soot formation.[36] Lieb's theorem is generally applicable to any bipartite system at half filling described by a repulsive Hubbard model, thus accounting for the magnetic ground states of many PCHs. However, magnetic correlations in non-benzenoid molecules, where the bipartite symmetry of the underlying lattice is broken, cannot be captured by this rule. This opens new opportunities to engineer spin correlations with non-benzenoid rings beyond the constraints of Lieb's theorem.[37,38] Among the existing non-benzenoid open-shell systems, most are reported to exhibit a singlet ($S = 0$) ground state, while systems with high-spin ground states are rare, with two notable examples being a tetrabenzo-Chichibabin hydrocarbon reported by Zeng et al.[23] and a cyclopenta-ring-fused oligophenylene decaradicaloid reported by Liu et al.,[27] with both systems exhibiting an open-shell triplet ($S = 1$) ground state. However, open-shell non-benzenoid systems with $S > 1$ remain elusive. Previously, there have been attempts to synthesize the non-benzenoid three-fold symmetric PCH truxene-5,10,15-triyl (**1**, Fig. 1b) in solution, which possibly exhibits an open-shell quartet ($S = 3/2$) ground state. Frantz et al.[39] attempted the formation of *mesityl*-substituted **1** via reduction of a triol precursor with SnCl₂ under acidic conditions, which, however, resulted in a dihydromonoradical product. Recently, Yang et al.[40] reported the synthesis of a derivative of **1** consisting of 1- and 9-anthracenyl substituents where delocalization of spin density to the anthracenyl substituents led to thermodynamic stabilization of the triradical, although a low spin concentration of less than one spin per molecule was found via solid state electron spin resonance measurements. However, isolation of the parent **1** hitherto remains elusive.

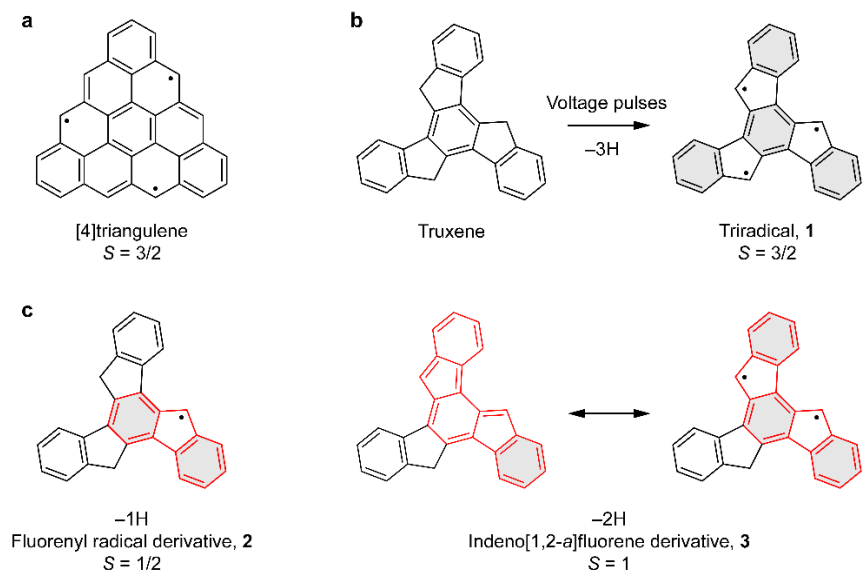

**Figure 1. On-surface reactions of truxene. (a)** Chemical structure of [4]triangulene. **(b)** Schematic representation of the on-surface generation of **1** by tip-induced dehydrogenation of truxene. **(c)** Reaction intermediates **2** and **3** observed during generation of **1** from truxene. Fluorenyl radical and indeno[1,2-*a*]fluorene moieties are highlighted in red. Gray filled rings denote Clar sextets.

Here, we report the synthesis of **1** and demonstrate its open-shell quartet ground state. Starting from truxene (10,15-dihydro-*5H*-diindeno[1,2-*a*;1′,2′-*c*]fluorene), an ambient-stable PCH, the tip of a combined scanning tunneling microscope (STM)/atomic force microscope (AFM) was used to sequentially cleave three C–H bonds, ultimately leading to the formation of the non-Kekulé species **1** on metal surfaces and thin insulating films. We study the electronic structure of **1** using a combination of scanning tunneling spectroscopy (STS) measurements and density functional theory (DFT) calculations, wherein we confirm the open-shell ground state of **1** and detect the orbital densities of the associated singly occupied molecular orbitals (SOMOs). Furthermore, our atomic manipulation-based synthetic approach allows us to isolate and image two reactive intermediates toward the synthesis of **1** (Fig. 1c), *viz.* derivatives of fluorenyl radical (**2**) and indeno[1,2-*a*]fluorene (**3**) – the indenofluorene regioisomer with the largest predicted diradical character and open-shell triplet ground state.

## RESULTS AND DISCUSSION

**Solution synthesis of truxene.** Truxene was obtained in 71% yield by trimerization of indan-1-one following a previously reported procedure by Dehmlow and Kelle.[41] Indan-1-one was added to a mixture of acetic acid and concentrated hydrochloric acid, and the solution was stirred for 16 h at 100 °C, and then poured on ice. The solid precipitate was washed with water, acetone and dichloromethane to give truxene as a white powder.

**Generation of triradical on coinage metal surfaces.** We started by exploring the formation of **1** through tip-induced dehydrogenation of truxene on Au(111) and Cu(111) surfaces. Dehydrogena-tion events were induced by positioning the STM tip over the center of the molecule and opening the feedback loop at typical tunneling conditions of $V = 0.2$ V and $I = 1.0$ pA (where $V$ and $I$ denote the sample bias and tunneling current, respectively). From this setpoint, the tip was retracted by about 5 Å to limit the tunneling current, and the sample bias was ramped to 4.0–4.6 V, where abrupt changes in the tunneling current indicated manipulation events (see Fig. S1 for a representative $I(V)$ curve for dehydrogenation of truxene). As shown in the sequence of constant-height AFM images in Fig. 2, dehydrogenation of truxene was reproducibly achieved on both Au(111) (Fig. 2a–d) and Cu(111) (Fig. 2e–h) to generate **1**. The adsorption conformation of **1** is qualitatively different on the two surfaces: in contrast to its largely planar adsorption on Cu(111) (Fig. 2g, h), **1** exhibits a non-planar adsorption conformation on Au(111), where the apical carbon atoms of the three pentagonal rings are not resolved in AFM imaging (Fig. 2d), explained by them being closer to the surface compared to the other carbon atoms of the molecule.[42] We tentatively attribute the difference in the adsorption conformation of **1** on the two surfaces to the different surface lattice parameters of Au(111) and Cu(111), such that commensurability of the apical carbon atoms of the pentagonal rings with the on-top atomic sites of the surface (which maximizes bonding interactions) is achieved for Au(111) but not for Cu(111). We note that based on the Clar structure of **1** (Fig. 1b), the pentagonal ring apices are expected to bear the largest spin densities, which likely makes them the most reactive moieties in **1**. Our interpretation is further supported by a previous work by Di Giovannantonio et al. on the synthesis of indeno[2,1–*b*]fluorene polymers on Au(111),[43] where both tilted and planar configurations of individual indenofluorene units were observed depending on whether or not the highest spin density-bearing carbon atoms were adsorbed on on-top Au sites.

**Electronic structure of triradical.** To experimentally access the electronic structure of **1**, we used two monolayer-thick (2ML) NaCl films on Cu(111) as an insulating layer to electronically decouple the molecule from the underlying metal surface. As shown in Fig. 3, formation of **1** on 2ML NaCl/Cu(111) could be reliably achieved, albeit larger sample biases (typically 5.4–5.7 V) at tip-sample distances of about 15 Å were required for dehydrogenation reactions than on Cu(111). On the 2ML NaCl/Cu(111) surface, the molecules were frequently displaced by interaction with the tip before any manipulation events occurred. Therefore, we studied molecules that were adsorbed at the edge of 3ML NaCl islands.[11,44]

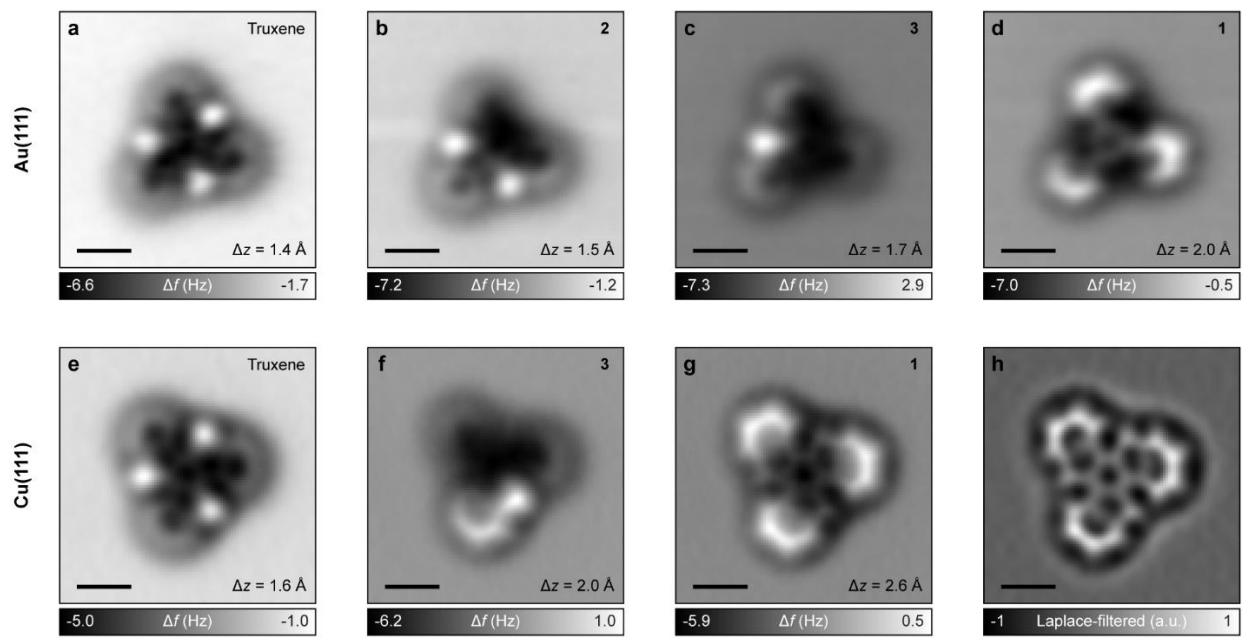

**Figure 2. Generation of 1 on coinage metal surfaces. (a–d)** Series of AFM images showing generation of **1** through tip-induced dissociation of individual hydrogen atoms from truxene on Au(111) – **(a)** truxene, where the three bright protrusions correspond to $CH_2$ moieties at the pentagonal ring apices, **(b)** intermediate **2, (c)** intermediate **3** and **(d)** triradical **1. (e–g)** Series of AFM images showing generation of **1** on Cu(111). **(h)** Laplace-filtered version of **(g)** revealing structural details of **1** (a.u. denotes arbitrary units). The tip height offset ($\Delta z$) for each panel is provided with respect to the initial STM setpoint of $V = 0.2$ V and, $I = 1.0$ pA **(a–d)** and 0.5 pA **(e–g)** above the respective bare metal surfaces. All scale bars: 5 Å.

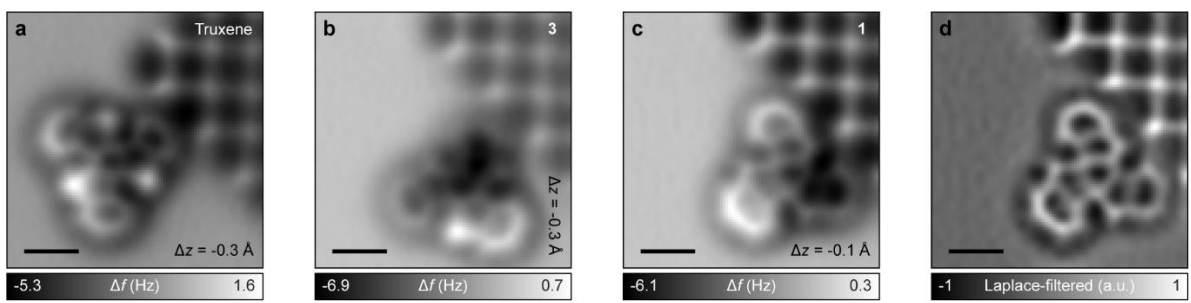

**Figure 3. Generation of 1 on NaCl. (a)** AFM image of truxene adsorbed on 2ML NaCl/Cu(111). The image frame also contains a 3ML NaCl patch in the upper right corner, where the bright and dark features correspond to $Cl^-$ and $Na^+$ ions, respectively. **(b, c)** Tip-induced dehydrogenation leading to the formation of **3 (b)** and **1 (c)**. **(d)** Laplace filtered version of **(c)**. STM setpoint for the AFM images: $V = 0.2$ V and $I = 0.5$ pA on 2ML NaCl. All scale bars: 5 Å.

To study the electronic structure of **1**, we first performed spin-polarized DFT calculations. Our calculations predict an open-shell quartet ground state of **1** in the gas phase, with the open-shell doublet ($S = 1/2$) state 100 meV higher in energy (see Fig. S2 for DFT calculations for **1** in the open-shell doublet state), implying a robust high-spin ground state of **1**. Fig. 4a shows the spin-polarized DFT energy spectrum of **1** around the first three molecular orbitals, which correspond to three SOMOs (denoted as $\Psi_1$, $\Psi_2$ and $\Psi_3$) with a ferromagnetic correlation between the populating spins. Furthermore, while $\Psi_2$ and $\Psi_3$ are degenerate, $\Psi_1$ is lower in energy for both the occupied (spin up) and unoccupied (spin down) channels. Also shown in Fig. 4b is the computed spin density distribution of **1**, where the largest spin density is present at the apices of the pentagonal rings, in agreement with the Clar structure of **1** (Fig. 1b).

Our STM experiments support the principal outcomes of our theoretical analyses. Differential conductance ($dI/dV$) spectrum acquired on **1** on 2ML NaCl/Cu(111) (Fig. 4c, d) reveals two distinct peaks at $V = -2.00$ V and $V = 1.15$ V, which we assign to the positive (PIR) and negative ion resonances (NIR), respectively. Figures 4e shows constant-height STM images acquired around the PIR, gap ($V = 0.20$ V) and NIR (see Fig. S3 for additional STM images). While the STM image acquired in the gap resembles the geometrical shape of **1**, images recorded at the PIR and NIR exhibit orbital densities where the salient features correspond to three prominent lobes at the apices of the pentagonal rings, along with three weaker lobes at the outer benzenoid rings.

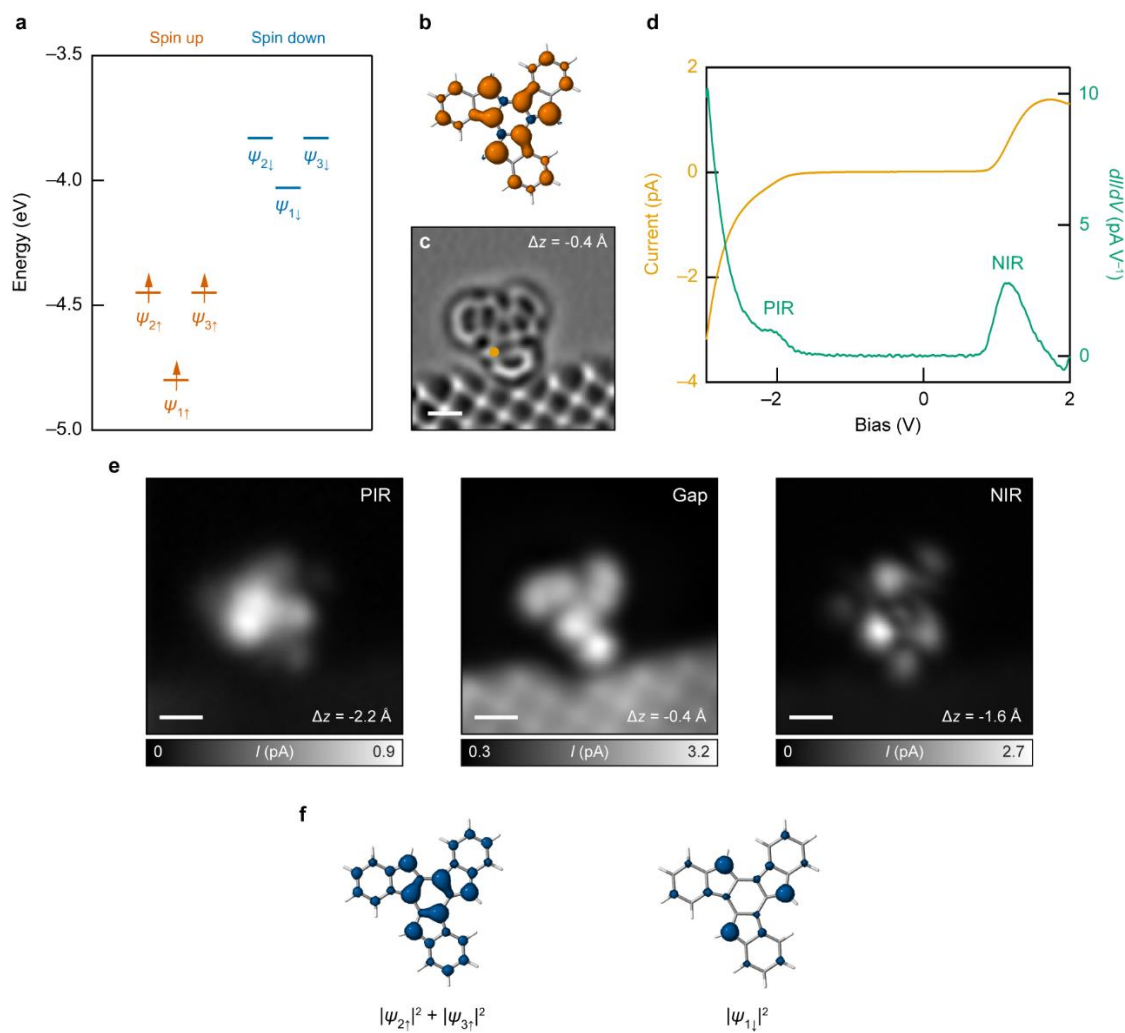

**Figure 4. Electronic characterization of 1. (a)** Open-shell quartet DFT energy spectrum of **1** in the gas phase. $\Psi_1$, $\Psi_2$ and $\Psi_3$ denote the SOMOs. Energies are referenced with respect to the vacuum level. **(b)** DFT computed spin density distribution of **1** in its quartet ground state. Orange and blue isosurfaces denote spin up and spin down densities, respectively (isovalue: $0.02e$ Å$^{-3}$). **(c)** Laplace-filtered AFM image of **1** on 2ML NaCl/Cu(111) (STM setpoint: $V = 0.2$ V and $I = 0.5$ pA on 2ML NaCl). **(d)** Constant-height $I(V)$ spectrum acquired on **1**, along with the corresponding $dI/dV$ spectrum obtained by numerical differentiation of the $I(V)$ curve (open feedback parameters: $V = -3.0$ V, $I = 4$ pA). Acquisition position is marked in (c) with a filled circle. **(e)** Constant-height STM images of **1** acquired around the PIR, gap and NIR with a CO-functionalized tip (open feedback parameters: $V = -1.8$ V (PIR), 0.2 V (gap) and 0.8 V (NIR), and $I = 0.5$ pA on 2ML NaCl). **(f)** DFT-LDOS maps of the frontier orbitals of **1** in the occupied (left) and unoccupied (right) channels for the quartet state (isovalue: $0.02e$ Å$^{-3}$). All scale bars: 5 Å.

Although **1** is an open-shell molecule, the PIR and NIR exhibit some differences. This is in contrast to other open-shell systems imaged with STM, where similar orbital densities at both bias polarities were observed due to electron tunneling to/from SO-MOs.[11,45] At the NIR, the lobes on the pentagonal rings of **1** appear rather localized with a weak orbital density at the center of the molecule. In contrast, the lobes at the pentagonal rings appear smeared out at the PIR, with significant orbital density at the molecular center. This difference may be rationalized by considering the DFT energy spectrum of **1** (Fig. 4a), where the frontier orbitals, which should dominate the contrast in STM imaging at the PIR and NIR, correspond to $\Psi_{2\uparrow}$ and $\Psi_{3\uparrow}$ in the occupied regime, and $\Psi_{1\downarrow}$ in the unoccupied regime. In Fig. 4f, we plot the DFT local density of states (LDOS) maps associated with these frontier orbitals, that is, $|\Psi_{1\downarrow}|^2$ and $|\Psi_{2\uparrow}|^2 + |\Psi_{3\uparrow}|^2$. For both maps, the predominant features correspond to the three lobes at the pentagonal

ring apices. However, in the occupied regime, a substantial electronic density is additionally present above the central benzenoid ring, which experimentally manifests at the PIR as the smearing out of the three lobes and increased orbital density at the molecular center.

Given the open-shell quartet ground state of **1**, magnetic excitations, *viz.* quartet-doublet spin excitations and/or Kondo resonances, are expected to be observed in **1**. On both Au(111) and Cu(111), we do not observe any signatures of magnetic excitations likely due to the strong electronic hybridization between **1** and the metal surfaces. Additionally, the high mobility of **1** on 2ML NaCl/Cu(111) precludes the employment of larger tunneling currents that are used to detect inelastic processes such as spin excitations.

**Isolation of reactive intermediate species.** Our atomic manipulation-based synthetic route also allowed us to isolate two intermediates toward the synthesis of **1**. Dissociation of a single hydrogen atom from truxene generates the non-Kekulé *S* = 1/2 fluorenyl radical moiety (**2**, Fig. 5a, b). Further cleavage of a second hydrogen atom from **2** leads to the generation of the indeno[1,2–*a*]fluorene moiety (**3**, Fig. 5c, d). Indeno[1,2–*a*]fluorene belongs to the general class of compounds known as indenofluorenes (Fig. 5e), which are non-benzenoid PCHs containing a conjugated array of alternating six- and five-membered rings. The presence of 20 π-electrons in indenofluorenes makes them formally antiaromatic, while presence of quinodimethane moieties in the conjugated framework endows them with potential open-shell character.[33] Among the five indenofluorene regioisomers shown in Fig. 5e, solution syntheses of compounds **I–IV** have been achieved in solution,[24,46–48] while **II–IV** have also been synthesized on surfaces.[43,49,50] Compound **V**, whose derivative **3** is achieved in the present work, is predicted to be the indenofluorene regioisomer with the largest open-shell diradical character.[33,43,51] The transient isolation of a derivative of **V** has only once been reported in solution by Dressler et al.,[51] who obtained structural proof of the stable dianion of **V**. Our work reports the direct generation of a derivative of **V**. In Fig. S4, we provide a detailed STS characterization of **3**. Briefly, **3** exhibits a frontier orbital gap of 3.25 eV on 2 ML NaCl/Cu(111), which is 0.10 eV larger than the gap of 3.15 eV for **1**, consistent with the increased quantum confinement in **3**. While our DFT calculations, in consistence with previous works,[51] predict an open-shell triplet ground state of **3** in the gas phase; based on the correspondence between our experimentally measured orbital densities and DFT-LDOS maps of the frontier orbitals of **3**, we presently cannot unambiguously assign the electronic ground state of **3** to either open- or closed-shell.

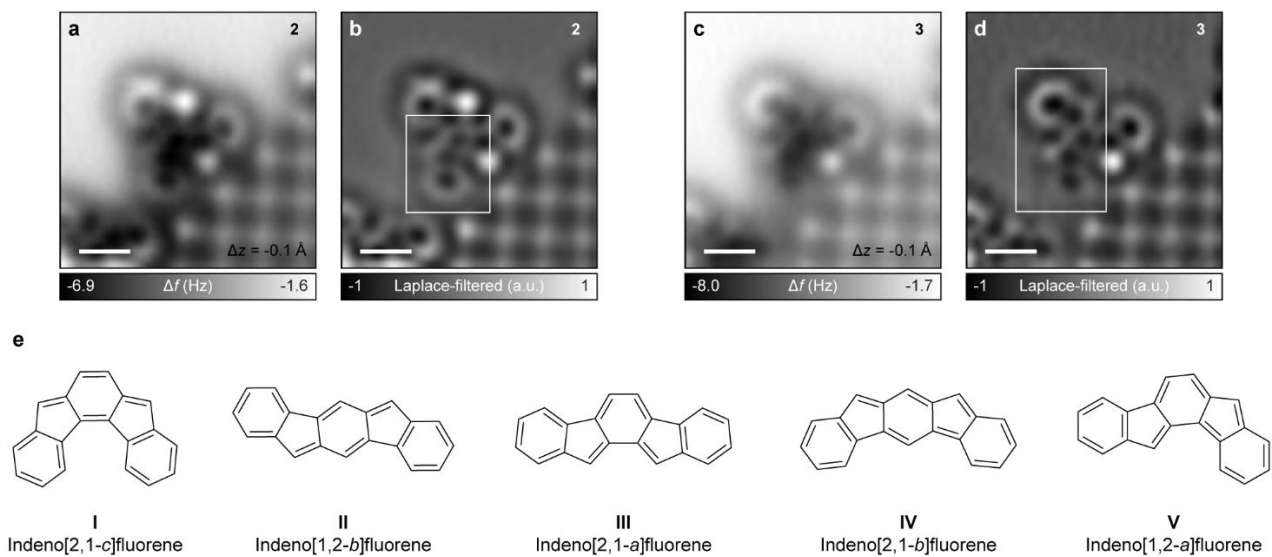

**Figure 5. Imaging of reaction intermediates. (a, b)** AFM image **(a)** and corresponding Laplace-filtered version **(b)** of intermediate **2**. The fluorenyl radical moiety is highlighted with a rectangle. **(c, d)** AFM image **(c)** and corresponding Laplace-filtered version **(d)** of intermediate **3**. The indeno[1,2–*a*]fluorene moiety is highlighted with a rectangle. STM setpoint for the AFM images: *V* = 0.2 V and *I* = 0.5 pA on 2ML NaCl. **(e)** Closed-shell Kekulé structures of the five indenofluorene regioisomers **I–V**. All scale bars: 5 Å.

## CONCLUSIONS

We have described the synthesis of a non-benzenoid spin-quartet PCH truxene-5,10,15-triyl (**1**) adsorbed on coinage metal and insulator surfaces. Our synthetic protocol combines in-solution synthesis of the stable hydrocarbon precursor truxene and its STM tip-induced dehydrogenation on surfaces to generate **1**. Our STS measurements, in combination with DFT calculations, demon-strate the open-shell ground state of **1**. The strong intramolecular ferromagnetic exchange (estimated to be 0.10 eV from DFT calculations) in combination with its three-fold symmetry, renders **1** a promising molecular building block toward fabrication of spin chains and lattices. Such structures, for example, may be fabricated on metal surfaces via installation of labile halogen atoms at the outer benzenoid rings of truxene, which should afford dehalogenative aryl-aryl coupling upon thermal annealing, or alternatively, they may also be fabricated on insulators via STM-based atom manipulation.[52] On a general note, our work presents a rare instance of syntheses of high-spin PCHs with non-benzenoid ring topologies.

## METHODS

**Sample preparation and scanning probe measurements.** STM and AFM measurements were performed with a home-built low-temperature STM/AFM system operating at a temperature of 5 K and base pressure below 1×10⁻¹⁰ mbar. Au(111) and Cu(111) single crystal surfaces were prepared by cycles of sputtering with Ne⁺ ions and annealing to 770 K. NaCl was evaporated onto Cu(111) surface held between 273–293 K, which leads to the growth of defect-free and predominantly bilayer NaCl(100) islands. Submonolayer coverages of truxene and CO molecules were deposited at a sample temperature of 10 K. STM images were acquired both in constant-current and constant-height modes, while *dI/dV* spectra were acquired in constant-height mode. Positive and negative values of the tip height offset Δ*z* denote tip approach and retraction from the STM setpoint, respectively. Bias voltages are provided with respect to the sample. Unless otherwise noted, all STM and STS measurements were performed with metallic tips. Non-contact AFM measurements were performed with a qPlus sensor[53] operated in frequency modulation mode,[54] with the oscillation amplitude kept constant at 0.5 Å. All AFM measurements were performed in constant-height mode with CO-functionalized tips.[5] STM and AFM images, and STS curves, were post-processed using Gaussian low-pass filters.

**DFT calculations.** DFT was employed using the FHI-aims code.[55] All molecules were independently investigated in the gas phase. The geometries were optimized with the *really tight* basis defaults. The Perdew-Burke-Ernzerhof[56] exchange-correlation functional, along with the van der Waals scheme by Tkatchenko and Scheffler,[57] were employed for structural relaxation. The convergence criteria were set to $10^{-3}$ eV Å$^{-1}$ for the total forces and $10^{-5}$ eV for the total energies.

## ASSOCIATED CONTENT

### Supporting Information

Representative $I(V)$ curve acquired during dehydrogenation of truxene, additional DFT calculations on **1** and **3**, additional STM data on **1** and STS measurements on **3** (PDF)

## AUTHOR INFORMATION

### Corresponding Author

*Shantanu Mishra (shm@zurich.ibm.com)
*Diego Peña (diego.pena@usc.es)
*Leo Gross (lgr@zurich.ibm.com)

### Author Contributions

S.M., Sh.F., K.K. and L.G. performed on-surface synthesis and scanning probe measurements. Sa.F and D.P. synthesized and characterized the precursor in solution. Sh.F. performed the DFT calculations. S.M. performed the tight-binding calculations, analyzed the data, and drafted the first version of the manuscript. All authors discussed the results and contributed to writing the manuscript.

### Notes

The authors declare no competing financial interests.

## ACKNOWLEDGMENT

We thank R. Allenspach for discussions. This work was financially supported by the European Union project SPRING (grant number 863098), the European Research Council Synergy grant Mol-DAM (grant number 951519), the Spanish Agencia Estatal de Investigación (PID2019-107338RB-C62 and PCI2019-111933-2), Xunta de Galicia (Centro de Investigación de Galicia accreditation 2019–2022, ED431G 2019/03) and the European Regional Development Fund.

# Supporting Information

# Non-benzenoid high-spin polycyclic hydrocarbons generated by atom manipulation

Shantanu Mishra,[1] Shadi Fatayer,[1] Saleta Fernández,[2]
Katharina Kaiser,[1] Diego Peña[2] and Leo Gross[1]

[1]IBM Research–Zurich, 8803 Rüschlikon, Switzerland

[2]Centro Singular de Investigación en Química Biolóxica e Materiais Moleculares (CiQUS) and Departamento de Química Orgánica, Universidade de Santiago de Compostela, 15782 Santiago de Compostela, Spain

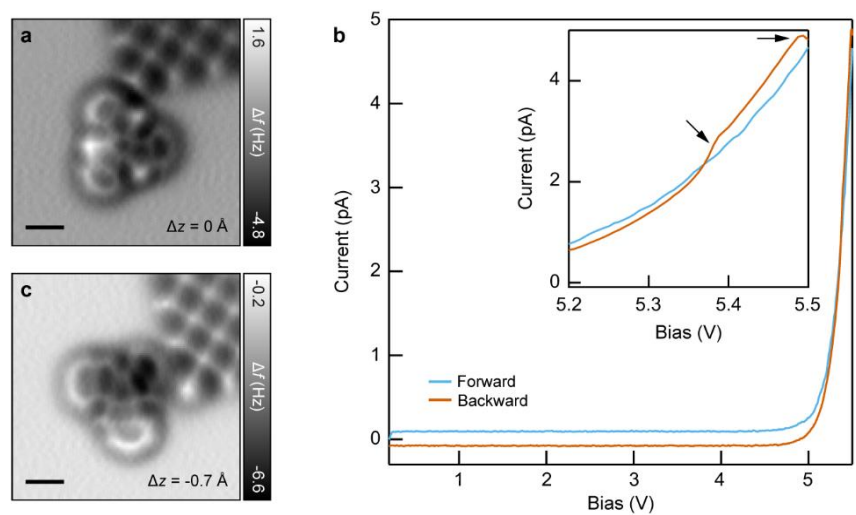

**Figure S1. Dehydrogenation of truxene on 2 ML NaCl/Cu(111). (a)** AFM image of truxene adsorbed on 2ML NaCl/Cu(111). The upper right corner of the image frame contains a 3ML NaCl patch. **(b)** $I(V)$ spectrum acquired during tip-induced dehydrogenation of truxene. To perform the dehydrogenation, the tip was relocated under constant-current conditions to the center of the molecule at a setpoint of $V$ = 0.2 V and $I$ = 0.5 pA. Thereafter, the feedback loop was switched off and the tip was retracted by 17 Å. Finally, the bias was ramped from 0.2 V to 5.5 V and backward at a constant sweep rate within 20 seconds. Blue and orange curves denote forward and backward $I(V)$ ramps, respectively. A small difference in the current of ~0.1 pA between forward and backward ramps, and the consequent difference to $I$ = 0 A for $V$ < 4.5 V results from capacitive coupling of the voltage ramp to the measured current signal. As shown in the zoom-in image in the inset, two jumps in the backward ramp can be discerned close to 5.5 V and 5.4 V (marked with arrows). **(c)** AFM image acquired after the bias ramp reveals generation of **1**. Note that the molecule was displaced from its original position after the manipulation event. STM setpoint for the AFM images: $V$ = 0.2 V and $I$ = 0.5 pA on 3ML **(a)** and 2ML **(c)** NaCl. All scale bars: 5 Å.

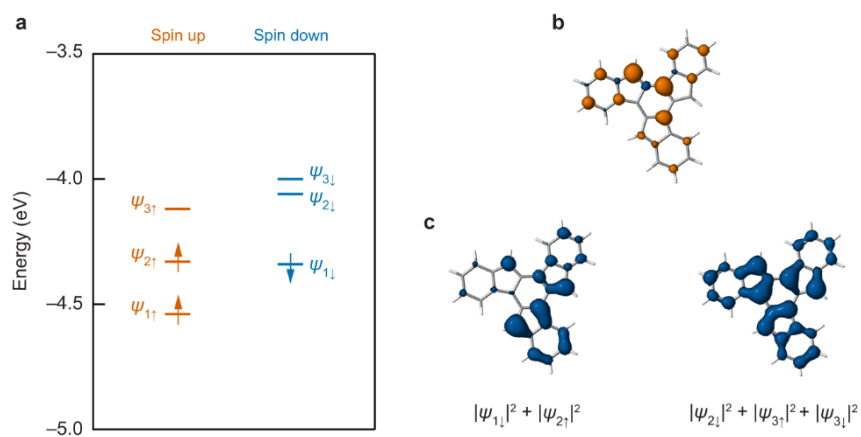

**Figure S2. DFT calculations for the open-shell doublet state of 1. (a)** Open-shell doublet DFT energy spectrum of **1** in the gas phase. $\Psi_1$, $\Psi_2$ and $\Psi_3$ denote the frontier orbitals. Energies are referenced with respect to the vacuum level. **(b)** DFT computed spin density distribution of **1** in its doublet state. Orange and blue isosurfaces denote spin up and spin down densities, respectively (isovalue: 0.02$e$ Å$^{-3}$). **(c)** DFT-LDOS maps of the frontier orbitals of **1** in the occupied (left) and unoccupied (right) channels for the doublet state (isovalue: 0.02$e$ Å$^{-3}$).

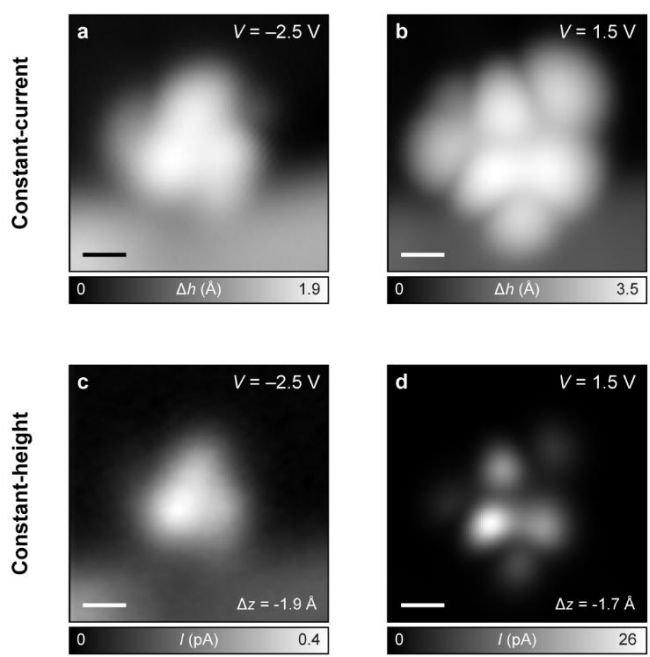

**Figure S3. Additional STM images of 1.** Constant-current **(a, b)** and constant-height **(c, d)** STM images of **1** acquired at $V = -2.50$ V **(a, c)** and 1.50 V **(b, d)**. The molecule corresponds to the same one shown in Fig. 4 of the main text. These biases correspond to values away from the resonance peaks of $-2.00$ V (PIR) and 1.15 V (NIR). Setpoints for constant-current imaging: $V = -2.50$ V **(a)** and 1.50 V **(b)**, $I = 1.0$ pA. $\Delta h$ denotes the apparent height. Open feedback parameters for constant-height imaging: $V = -2.50$ V **(c)** and 1.50 V **(d)**, $I = 0.5$ pA. All scale bars: 5 Å.

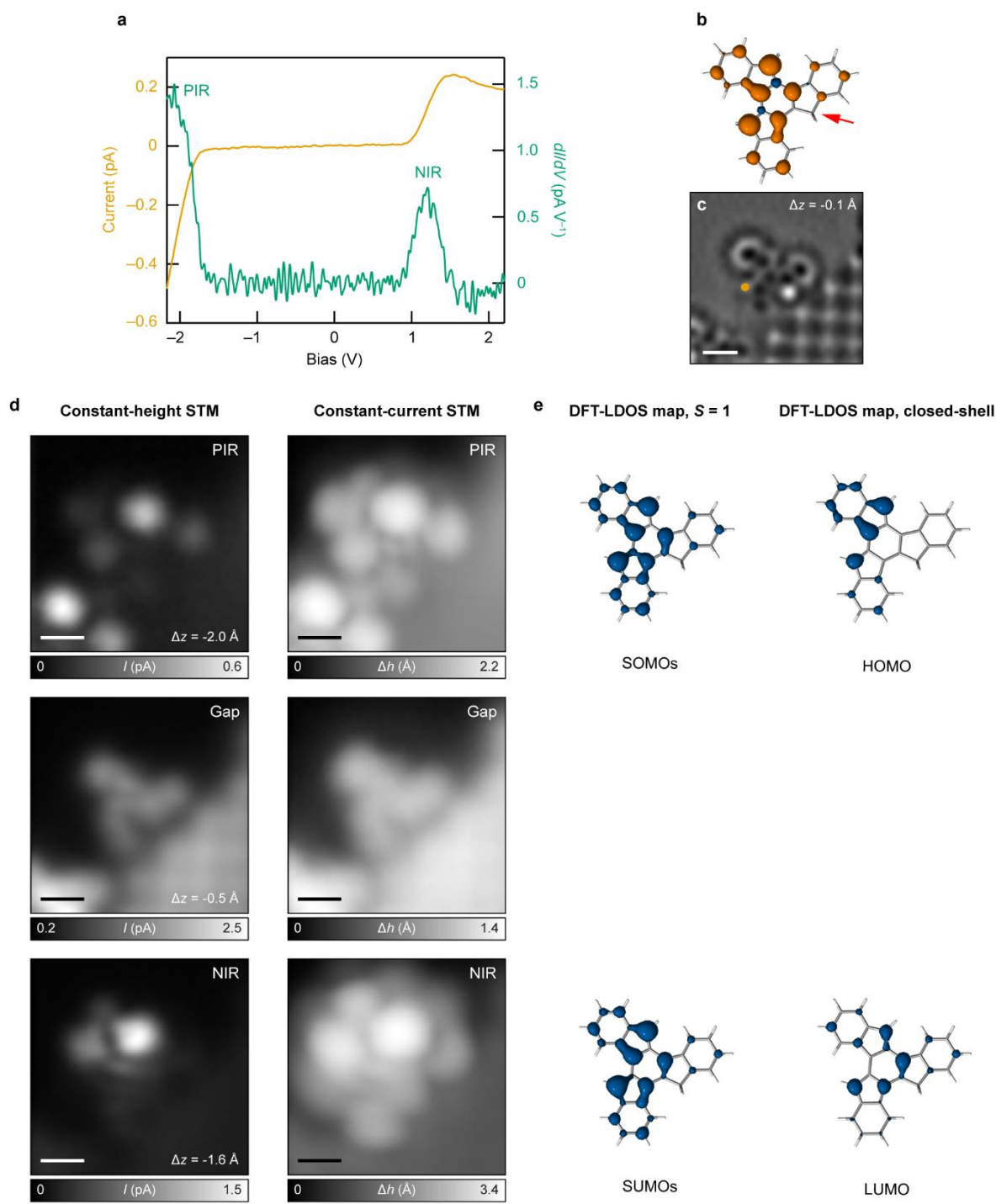

**Figure S4. Electronic characterization of 3. (a)** Constant-height $I(V)$ spectrum acquired on **3,** along with the corresponding $dI/dV$ spectrum obtained by numerical differentiation of the $I(V)$ curve (open feedback parameters: $V$ = −2.2 V, $I$ = 0.5 pA). Acquisition position is marked in **(c)** with a filled circle. Positive and negative ion resonances are detected at −2.05 V and 1.20 V, respectively, leading to a frontier electronic gap of 3.25 eV. **(b)** DFT computed spin density distribution of **3** in its triplet ground state. The largest spin densities are present at the apices of the pentagonal rings of the indeno[1,2−$a$]fluorene moiety. Orange and blue isosurfaces denote spin up and spin down densities, respectively (isovalue: 0.02$e$ Å$^{-3}$). The arrow indicates the CH$_2$ moiety of **3**. **(c)** Laplace-filtered AFM image of **3** on 2ML NaCl/Cu(111), adsorbed next to a 3ML NaCl patch (STM setpoint: $V$ = 0.2 V and $I$ = 0.5 pA on 2ML NaCl). The image is also shown in Fig. 5. **(d)** Constant-height (left column) and constant-current (right column) STM images of **3** around the PIR (top row), gap (middle row) and NIR (bottom row), acquired with a CO-functionalized

tip. Open feedback parameters for constant-height imaging: $V$ = −1.8 V (PIR), 0.2 V (gap) and 0.9 V (NIR), and $I$ = 0.5 pA on 2ML NaCl. Setpoints for constant-current imaging: $V$ = −1.8 V (PIR), 0.2 V (gap) and 1.1 V (NIR), and $I$ = 0.5 pA. **(e)** DFT-LDOS maps of the frontier orbitals of **3** for the open-shell triplet (left column) and closed-shell (right column) states (isovalue: 0.02$e$ Å$^{-3}$). For the open-shell triplet state, the frontier orbitals consist of two degenerate singly occupied molecular orbitals, and as expected, they exhibit similar appearances in both the occupied (denoted SOMOs, containing the majority spin) and the corresponding unoccupied (denoted SUMOs, containing the minority spin) channels. In contrast, the frontier orbitals for the closed-shell solution correspond to the hybridized highest occupied and lowest unoccupied molecular orbitals (HOMO and LUMO, respectively), which appear different. Based on the correspondence between our experimental data and the DFT-LDOS maps of **3**, we cannot unambiguously assign the ground state of **3** in the present case to either open- or closed-shell. All scale bars: 5 Å.